\documentstyle[12pt]{article}
\begin{document}
\begin{titlepage}
\title{Autoparallel Orbits in Kerr Brans-Dicke Spacetimes}
\author{ H. Cebeci\footnote{E.mail: hcebeci@metu.edu.tr}\\{\small Department of Physics, Middle East Technical University}\\{\small 06531 Ankara, Turkey}\\  \\
T. Dereli\footnote{E.mail: tdereli@ku.edu.tr}\\{\small Department of Physics,  Ko\c{c} University}\\{\small 80910 Sar{\i}yer,  \.{I}stanbul, Turkey} \\     \\
R. W. Tucker\footnote{E.mail: r.tucker@lancaster.ac.uk} \\
{\small Department of Physics, Lancaster University}\\ {\small
Lancaster LA1 4YB, UK}}
\date{  }

\maketitle

%\bigskip

%\bigskip

\begin{abstract}
\noindent The bounded orbital motion of a massive spinless test
particle in the background of a Kerr Brans-Dicke geometry is
analysed in terms of worldlines that are auto-parallels of
different metric compatible  space-time connections. In one case
the connection is that of Levi-Civita with zero-torsion. In the
second case the connection has torsion determined by the gradient
of the Brans-Dicke background scalar field. The calculations
permit one in principle to discriminate between these
possibilities.
\end{abstract}
\end{titlepage}

\bigskip

\section{Introduction}

\bigskip

 In  general relativity  it is commonly assumed
that idealised massive spinless test {\it particles} have
spacetime histories that coincide with time-like geodesics
associated with the spacetime metric when acted on by gravitation
alone.
 Such assumptions
embody such notions as the ``equivalence principle'', the
``equality of inertial and gravitational mass" and the tenets of
``special relativity".   Such a geometrical description of
gravitational interactions is formulated in terms of Einstein's
torsion-free metric-compatible (affine) connection. The  particle
histories then have self-parallel 4-velocities (tangent vectors)
and may be termed \lq\lq Levi-Civita auto-parallels\rq\rq.
If a massive test particle moves in a bounded orbit in the
geometry described by the exterior Schwarzshild metric  one may
use such hypotheses to calculate its orbital perihelion shift per
revolution.  Despite competing perturbations, such a perihelion
shift of the planet Mercury  in the gravitational field of the Sun
is regarded as one of the classical tests of Einstein's theory of
gravitation \cite{will}, \cite{will2}. Further evidence for these
hypotheses is sought from observations of the pulse rate of the
binary pulsar PSR 1913+16 that appears to be speeding up due to
gravitational radiation \cite{pulsar}.
A number of efforts have been made \cite{einstein},
\cite{papapetrou}, \cite{dixon} to prove the {\it geodesic
hypothesis} for test particles from a field theory approach but
none are entirely convincing. However, (for spinless test
particles)  this hypothesis  has almost become elevated to one of
the natural laws of physics \cite{weinberg}.

In 1961 Brans and Dicke \cite{brans}  suggested a modification of
Einsteinian gravitation by introducing a single additional scalar
field with particular gravitational couplings to matter via the
spacetime metric. They suggested that the experimental detection
of such a new scalar component to gravitation might have been
overlooked in traditional tests of gravitational theories. Their
theory is arguably the simplest modification to Einstein's
original description. Since then a number of astrophysical
observations have indicated that gravitational scalar fields may
indeed have relevance to the dynamics of matter.
 Furthermore on the theoretical side, modern
``low energy effective string theories" are replete with scalar
fields and most unified theories of the strong and electroweak
interactions predict such fields with astrophysical implications
\cite{brans97}.

Although the gravitational field equations were modified by Brans
and Dicke \cite{brans}, they  still assumed that the motion of
test particles under the influence of gravitation  was described
by a Levi-Civita auto-parallel associated with the metric derived
from the Brans-Dicke field equations. Dirac later suggested
\cite{Dirac} that  it was more natural to generate the motion of a
test particle from a Weyl invariant action principle and that such
a motion in general differed from a Brans-Dicke Levi-Civita
auto-parallel. Although Dirac was concerned with the
identification of electromagnetism with aspects of Weyl geometry,
even for neutral test particles it turns out that such test
particles would follow auto-parallels of a connection with
torsion. In Ref. \cite{DT} we have shown that the theory of Brans
and Dicke  can be reformulated as a field theory on a spacetime
with dynamic torsion $T$ determined by the gradient of the
Brans-Dicke scalar field $\Phi$:
\begin{equation}
 T = e^a \otimes \frac{d\Phi}{2\Phi}\otimes X_a  - \frac{d\Phi}{2\Phi}
\otimes {e^a}\otimes X_a\label{tor}
\end{equation}
in any coframe $\{e^a\}$ with dual $\{X_a\}$.
 Although (in the absence of spinor fields) no new physics {\it of
the fields} can arise from such a reformulation, the behaviour of
spinless {\it massive test particles} in such a geometry with
torsion is less clear cut.   Two natural  alternatives present
themselves. One may assert  that their histories are {\sl either}
time-like geodesics associated with auto-parallels of the
Levi-Civita connection (as did Brans and Dicke) {\sl or}  the
auto-parallels of the non-Riemannian connection with torsion. In
\cite{dereli1, dereli2} we have shown that it is possible to
theoretically compare these alternatives for the history of a mass
in orbit about a spherically symmetric source of scalar-tensor
gravity by regarding it as a spinless test particle. In principle
such calculations could be confronted with observation in space
experiments that measure the orbital parameters in binary systems.
In practice there are many competing perturbations that may
contribute significantly. In particular a rotating gravitational
source will be expected to produce a modification whose magnitude
will depend on both its mass and angular momentum as well as the
scalar field. It is of interest therefore to compare the
auto-parallel orbits derived from the two connections above for a
spinless test particle in the metric of a spinning source.

In this note we perform such a calculation for a particle in the
geometry derived from the exact Kerr Brans-Dicke metric and
scalar field found in \cite{intosh} (See also \cite{kim}). This solution to the
Brans-Dicke equations is asymptotically flat and reduces to the
Kerr metric solution to Einstein's equations when the scalar
field is replaced by a constant. The solution describes a
stationary and axially symmetric metric and depends on parameters
that may be identified with the scalar charge, mass and angular
momentum of a localised source.

%%%%%%%%%%%%%%%%%%%%%%%%%%%%%%%%%%%%%%%%%%%%%%%%%%%%%%%%%%%
\bigskip

\section{The motion of massive test particles}
\bigskip

The Kerr Brans-Dicke solution found in \cite{intosh} can be
written in  Boyer-Lindquist coordinates $(t, r, \theta, \varphi)$
as;
\begin{eqnarray}
g &=& \Phi _{0}^{-1}\left(
\frac{r-(M+\sqrt{M^{2}-l^{2})}}{r-(M-\sqrt{M^{2}-l^{2})}}\right) ^{-\frac{1}{2}A}\{-\frac{\Sigma \Delta }{P}dt^{2}+%
\frac{P\sin ^{2}\theta }{\Sigma }(d\varphi -\frac{2Mlr}{P}dt)^{2} \nonumber \\
 & & + \Sigma \left( \frac{(r-M)^{2}-(M^{2}-l^{2})}{(r-M)^{2}-(M^{2}-l^{2})\cos
^{2}\theta }\right) ^{2kA^{2}}(d\theta ^{2}+\frac{dr^{2}}{\Delta
})\}
\end{eqnarray}
with the scalar field,
\begin{equation}
\Phi=\Phi_{0}
\left(\frac{r-(M+\sqrt{M^2-l^2})}{r-(M-\sqrt{M^2-l^2})}\right)^{\frac{A}{2}}
\end{equation}
where $M > l $ denotes the source mass  and $l$  its  angular
momentum
 per unit mass. The constant $A$ determines the strength of the scalar field
  and $k = \omega + 3/2$, in terms of the    Brans-Dicke parameter
$\omega$. As usual
\begin{equation}
\begin{array}{c}
\Sigma =r^{2}+l^{2}\cos ^{2}\theta \\
\Delta =r^{2}+l^{2}-2Mr \\
P=\Delta \Sigma +2Mr(r^{2}+l^{2}).
\end{array}
\end{equation}

We first examine test particle orbits $C$ for massive spinless
particles that follow Levi-Civita autoparallels. The equations of
motion are
$$
\hat {\nabla}_{\dot C} \dot C = 0
$$
in terms of  the torsion-free Levi-Civita connection $ \hat
{\nabla}$ and 4-velocity $\dot C$ normalised according to
\begin{equation}
\label{AA} {\bf g}(\dot C,\dot C) = -1 ,
\end{equation}
(throughout units are adopted such that $G=1$ and  $c=1$). If $C:
\tau \mapsto x^\mu(\tau)$  in terms of proper time $\tau$ these
equations yield
\begin{equation}
\frac{d}{d\tau}\left(\frac {dx^{\mu}}{d\tau} \right)+\{
{}^{\mu}_{\nu \lambda}\} \frac{dx^{\nu}}{d\tau}
\frac{dx^{\lambda}}{d\tau} = 0.
\end{equation}
The  metric  above has two independent Killing vectors
$\partial_{t}$ and $\partial_{\varphi}$. These  generate two
constants of motion: the  particle energy and orbital angular
momentum. Since the orbits are planar we take $\theta = \frac
{\pi}{2}$  and set
\begin{equation}
\bar{E} = m ( \dot {t} g_{tt} + \dot {\varphi} g_{t \varphi} ) ,
\label{BB}
\end{equation}
\begin{equation}
\bar{L} = m ( \dot {t} g_{t \varphi} + \dot {\varphi} g_{\varphi
\varphi}) . \label{CC}
\end{equation}
One can express $\dot r$ in (\ref{AA}) in terms of $\dot t$ and
$\dot \varphi$ and metric components on the orbit. Eliminating
$\dot t$ and $\dot \varphi$ from equations (\ref{BB}) and
(\ref{CC}) and substituting into (\ref{AA}),  the orbit equation
for the particle equation may be written:
\begin{equation}
\left(\frac{dr}{d\varphi}\right)^{2}=\frac{-\Delta
\Phi^{-2}}{(g_{t \varphi} \tilde{E}-g_{tt} \tilde{L})^{2} g_{rr}}
\left\{\Phi^{-2}\Phi_{0} \Delta + 2g_{t\varphi} \tilde{E}
\tilde{L} - g_{tt} \tilde{L}^{2} - g_{\varphi \varphi}
\tilde{E}^{2} \right\}
\end{equation}
where $\tilde{L}=\frac{\bar{L} (\Phi_{0})^{1/2}}{m}$ and
$\tilde{E}=\frac{\bar{E} (\Phi_{0})^{1/2}}{m}$. Now define,

%\begin{equation}
%\begin{array}{c}
%A(r)=\frac{4 M^{2} l^{2} - \Delta r^{2}}{P_{1}} \quad B(r)=\frac {-2 M l}{r}  \\
%C(r)=\frac{P_{1}}{r^{2}}
%\quad \Phi_{1}(r)=\left(\frac{r-(M+\sqrt{M^{2}-l^{2}})}{r-(M-\sqrt{M^{2}-l^{2}})}\right)^{-\frac{A}{2}} \\
%P_{1}(r)=(r^{2}+l^{2})r^{2}+2Ml^{2}r \quad
%G(r)=\frac{r^{2}}{\Delta} \left(\frac{r^{2}+l^{2}-2 M r }{r^{2}+
%M^{2} - 2 M r } \right)^{2 k A^{2}}
%\end{array}
%\end{equation}

\begin{equation}
\begin{array}{c}
A(r)=\frac{4 M^{2} l^{2} - \Delta r^{2}}{P_{1}}  {-2 M l}{r}
\end{array}
\end{equation}

\begin{equation}
\begin{array}{c}
 B(r)=\frac {-2 M l}{r}
\end{array}
\end{equation}

\begin{equation}
\begin{array}{c}
C(r)=\frac{P_{1}}{r^{2}}
\end{array}
\end{equation}

\begin{equation}
\begin{array}{c}
\Phi_{1}(r)=\left(\frac{r-(M+\sqrt{M^{2}-l^{2}})}{r-(M-\sqrt{M^{2}-l^{2}})}\right)^{-\frac{A}{2}}
\end{array}
\end{equation}

\begin{equation}
\begin{array}{c}
P_{1}(r)=(r^{2}+l^{2})r^{2}+2Ml^{2}r
\end{array}
\end{equation}

\begin{equation}
\begin{array}{c}
 G(r)=\frac{r^{2}}{\Delta}
\left(\frac{r^{2}+l^{2}-2 M r }{r^{2}+ M^{2} - 2 M r } \right)^{2
k A^{2}}
\end{array}
\end{equation}
and introduce  $u=\frac{1}{r}$ so that:
\begin{eqnarray}
\left(\frac{du}{d\varphi}\right)^{2}&=&\frac{-u^{4}\Delta(1/u)}
{\left(B(1/u)\tilde{E}-\tilde{L}A(1/u)\right)^{2}G(1/u)}
 \{\Phi_{1}(1/u) \Delta(1/u) + 2 B(1/u) \tilde{E} \tilde{L} - \nonumber \\
& & C(1/u) \tilde{E}^{2} - \tilde{L}^{2} A(1/u) \}.
\end{eqnarray}

To analyse this equation we  employ the physically motivated
approximations discussed in \cite{dereli2}. Thus if the radius of
a (weak field)  Newtonian orbit is much larger than the
corresponding Schwarzschild radius of the source, one may expand
this equation around $u=0$ up to third order in order to compare
its solutions with those in a  Schwarzschild background:

\begin{equation}
\left(\frac{du}{d\varphi}\right)^{2} \simeq
S_{0}+S_{1}u+S_{2}u^{2}+S_{3}u^{3}
\end{equation}
where the constants:
\begin{equation}
S_{0} = \frac{1}{\tilde{L}^{2}}(\tilde{E}^{2}-1) ,
\end{equation}
\begin{equation}
S_{1} =
\frac{1}{\tilde{L}^{3}}4Ml\tilde{E}(\tilde{E}^{2}-1)+\frac{1}{\tilde{L}
^{2}}(2M-\sqrt{M^{2}-l^{2}}A) ,
\end{equation}
\begin{eqnarray}
S_{2}
&=&-1+\frac{1}{\tilde{L}^{2}}\{-\frac{1}{2}(M^{2}-l^{2})A^{2}+3l^{2}
(\tilde{E}^{2}-1) \nonumber \\
 & & + M\sqrt{M^{2}-l^{2}}A
+2k(M^{2}-l^{2})(\tilde{E}^{2}-1)A^{2}\} \nonumber \\
 & & + \frac{1}{\tilde{L}^{3}}\{8M^{2}
\tilde{E}^{3}l -4Ml\tilde{E}\sqrt{M^{2}-l^{2}}A\}  \nonumber \\
 & &+\frac{12M^{2}l^{2}\tilde{E}^{2}}{\tilde{L}^{4}}(\tilde{E}^{2}-1) ,
\end{eqnarray}
\begin{eqnarray}
S_{3} &=&
2M+\frac{1}{\tilde{L}^{2}}\{-3l^{2}\sqrt{M^{2}-l^{2}}A-\frac{1}{3}
\sqrt{M^{2}-l^{2}}(4M^{2}-l^{2})A \nonumber \\
& & + 6Ml^{2}\tilde{E}^{2}
+2M^{2}\sqrt{M^{2}-l^{2}}A+4k\tilde{E}^{2}M(M^{2}-l^{2})A^{2} \nonumber \\
 & & - (2k+\frac{1}{6})
\left( M^{2}-l^{2}\right) ^{3/2}A^{3}\}
+\frac{1}{\tilde{L}^{3}}\{-2Ml\tilde{E}(M^{2}-l^{2})A^{2} \nonumber \\
& & -4M^{2}l\tilde{E}
\sqrt{M^{2}-l^{2}}A+16M^{3}l\tilde{E}^{3}  \nonumber \\
 & & + 8k\tilde{E}Ml(\tilde{E}^{2}-1)(M^{2}-l^{2})A^{2}+12M\tilde{E}l^{3}
(\tilde{E}^{2}-1)\} \nonumber \\
 & &
+\frac{12}{\tilde{L}^{4}}\{2M^{3}\tilde{E}^{2}l^{2}(2\tilde{E}^{2}-1)-M^{2}
\tilde{E}^{2}l^{2}\sqrt{M^{2}-l^{2}}A\} \nonumber \\
& & + \frac{1}{\tilde{L}^{5}}32l^{3}M^{3}
\tilde{E}^{3}(\tilde{E}^{2}-1) .
\end{eqnarray}
All the terms in $S_{2}$ except $-1$ and all the terms in $S_{3}$
 give corrections to the Newtonian orbital equation.

By contrast we now compare this equation with the one obtained by
assuming the worldline is a timelike  auto-parallel of a
particular connection with torsion:
$$
\nabla_{\dot C} \dot C = 0 .
$$
Here $\nabla$ denotes the connection with the torsion (\ref{tor})
specified by the scalar field in terms of the 2-forms
$$T^{a}=e^{a} \wedge
\frac{d\Phi}{2\Phi}$$
 and the  4-velocity $\dot C$ is again
normalized with
$$
{\bf g}(\dot C,\dot C) = -1 .
$$
It is possible to re-write the worldline equation in terms of the
Levi-Civita connection $\hat \nabla$ as
$$
\widetilde{\hat \nabla_{\dot C} \dot C}=-\frac{1}{2\Phi}
\iota_{\dot C} (d\Phi \wedge \tilde{\dot C})
$$
where for any vector field  $V$, $\tilde V={\bf g}(V,-)$ is the
metric related 1-form.
This may be  further simplified to
$$
\widetilde{\hat \nabla_{\dot C}(\Phi^{1/2} \dot C)}=-d\Phi^{1/2} ,
$$
restricted to $C$, which in local coordinates gives:
\begin{equation}
\frac{d}{d\tau}\left(\Phi^{1/2}\frac{dx^{\mu}}{d\tau}\right) +
\Phi^{1/2}\{ {}^{\mu}_{\nu \lambda} \}\frac{dx^{\nu}}{d\tau}
\frac{dx^{\lambda}}{d\tau}= -g^{\mu
\nu}\frac{\partial_{\nu}\Phi}{2\Phi^{1/2}} .
\end{equation}
For any Killing vector $K$ with $K\Phi = 0$, the expression
$$
\gamma_{K}=\Phi^{1/2}{\bf g}(K,\dot C)
$$
is constant along the worldline of the particle. As  before the
Killing vectors $K_t=\partial_t$ and
$K_{\varphi}=\partial_{\varphi}$ generate, respectively, two
constants of motion $E$ and $L$, corresponding to energy and
orbital angular momentum. Thus
\begin{equation}
E = m \left(\frac{\Phi}{\Phi_0}\right)^{1/2}( \dot {t} g_{tt} +
\dot {\varphi}g_{t \varphi} ) , \label{EE}
\end{equation}
\begin{equation}
L = m \left(\frac{\Phi}{\Phi_0}\right)^{1/2}( \dot {t} g_{t
\varphi} + \dot {\varphi} g_{\varphi \varphi} ) \label{BA}
\end{equation}
in terms of  metric functions evaluated on planar orbits.
Eliminating $\dot t$ and $\dot \varphi$ from equations (\ref{EE})
and (\ref{BA}) and substituting in (\ref{AA}), one obtains the new
auto-parallel orbit equation as
\begin{equation}
\left(\frac{dr}{d\varphi}\right)^{2}=\frac{-\Delta
\Phi^{-2}}{(g_{t \varphi} \hat{E}-g_{tt} \hat{L})^{2} g_{rr}}
\left\{\Phi^{-1} \Delta + 2g_{t\varphi} \hat{E} \hat{L} - g_{tt}
{\hat{L}}^{2} - g_{\varphi \varphi}{\hat{E}}^{2} \right\}
\end{equation}
where $\hat{E}=\frac{E (\Phi_0)^{1/2}}{m}$ and $\hat{L}=\frac{L
(\Phi_0)^{1/2}}{m}$. Expressed in terms of the variable
$u=\frac{1}{r}$, it becomes
\begin{eqnarray}
\left(\frac{du}{d \varphi}\right)^{2}&=&\frac{-u^{4}
\Delta(1/u)}{\left(B(1/u)\hat{E} - \hat{L} A(1/u) \right)^{2} G(1/u) }
\{ \Delta(1/u) + 2 B(1/u) \hat{E} \hat{L} - \nonumber \\
& & C(1/u) \hat{E}^{2} - \hat{L}^{2} A(1/u) \} .
\end{eqnarray}
Expanding  to  third order in $u$ as before, one finds
\begin{equation}
\left(\frac{du}{d\varphi}\right)^{2} \simeq C_{0} + C_{1} u +
C_{2} u^{2} +C_{3} u^{3}
\end{equation}
in terms of the constants:
\begin{equation}
C_{0} = \frac{1}{\hat{L}^{2}}(\hat{E}^{2}-1)
\end{equation}
\begin{equation}
C_{1} =
2\frac{M}{\hat{L}^{2}}+4\frac{M\hat{E}l}{\hat{L}^{3}}(\hat{E}^{2}-1)
\end{equation}
\begin{eqnarray}
C_{2} &=&
\frac{1}{\hat{L}^{2}}\{3l^{2}+2kA^{2}(M^{2}-l^{2})\}(\hat{E}^{2}-1)
+\frac{1}{\hat{L}^{3}}8M^{2}l\hat{E}^{3} \nonumber \\
 & & + \frac{1}{\hat{L}^{4}}12M^{2}l^{2}
\hat{E}^{2}(\hat{E}^{2}-1)-1
\end{eqnarray}
\begin{eqnarray}
C_{3} &=& 2M+\frac{1}{\hat{L}^{2}}\{(6M-4MkA^{2})l^{2}+4kM^{3}A^{2}\}\hat{E}^{2} \nonumber \\
 & & + \frac{1}{\hat{L}^{3}}\{[(12M-8kMA^{2})l^{3}+8kM^{3}lA^{2}]\hat{E}(\hat{E}^{2}
-1)+16M^{3}\hat{E}^{3}l\} \nonumber \\
 & & +
\frac{1}{\hat{L}^{4}}\{24M^{3}l^{2}\hat{E}^{2}(2\hat{E}^{2}-1)\}+\frac{1}
{\hat{L}^{5}}\{32M^{3}l^{3}\hat{E}^{3}(\hat{E}^{2}-1)\} .
\end{eqnarray}
The first three terms of $C_{2}$ and all terms in  $C_{3}$ imply
corrections to  Newtonian orbits.

We note that both orbit equations have been written in the form:
\begin{equation}
\left(\frac{du}{d\varphi}\right)^{2} \simeq g(u)=L_{0} + L_{1} u
+ L_{2} u^{2} + L_{3} u^{3}. \label{CA}
\end{equation}
so their solutions can be analysed in terms of the corresponding
constants according to the roots of the equation $g(u)=0$.
Suppose first that all three roots  are real, i.e.
$$
4 L_{1}^3 L_{3} + 4 L_{0} L_{2}^3 - L_{1}^2 L_{2}^2 + 27 L_{0}^2
L_{3}^2 - 18 L_{0} L_{1} L_{2} L_{3} < 0 .
$$
Suppose further that the roots are distinct and ordered to satisfy
$u_{1} < u_{2} < u_{3}$. Then
$$
u_{1}+u_{2}+u_{3}=-\frac{L_{2}}{L_{3}} .
$$
From the orbit equation,  $g(u)\geq 0$ throughout the motion.
Thus, $g(u)$ will have a local maximum between $u_{1}$ and
$u_{2}$. Hence, for a bounded  orbit, $u_{1}$ corresponds to
aphelion and $u_{2}$ corresponds to perihelion. Consider the
following cases \cite{schmidt}:
\bigskip

i. If $u_{1} > 0$, one obtains  bounded orbits of \lq\lq elliptic
\rq\rq type. This requires that both $L_{0}$ and $L_{2}$ be
negative provided that $L_{3}>0$. Then the particle is confined to
the interval $u_{1} < u < u_{2}$. (If $u_1 = u_2$,  one obtains
circular orbits.)

ii.  If $u_{1}=0$, one obtains open orbits of \lq\lq parabolic
\rq\rq type. This requires that $L_{0}=0$.  This is possible for
orbits associated with both Levi-Civita and torsional connections
provided $E^2 = \Phi^{-1}_{0} m^2$.

iii.  If $u_{1} < 0$, one obtains open orbits of \lq\lq
hyperbolic\rq\rq type. This requires that $E^2 > \Phi^{-1}_{0}
m^2$ provided that $L_{3}
> 0$ where $L_{3}=C_{3}$ if the orbit  is associated with an autoparallel
of the torsional connection
 and provided $L_{3}=S_{3}$ if it is associated with the Levi-Civita connection.

\bigskip
\newpage

\section{The analysis of bounded orbits}

\bigskip

\noindent We are interested here  in bounded orbits \cite{wilkins} in which case the general
solution of (\ref{CA}) can be
expressed in terms of Jacobian elliptic functions. By introducing
the variables
$$
x=\frac{1}{2} \varphi \sqrt{L_{3}\left(u_{3}-u_{1}\right)}, \qquad
y=\sqrt{\frac{u-u_{1}}{u_{2}-u_{1}}}
$$
(\ref{CA}) becomes
\begin{equation}
\left(\frac{dy}{dx}\right)^2 = \left(1-y^{2} \right)\left(1-p^2
y^2 \right) \label{EC}
\end{equation}
with $p=\sqrt{\frac{u_{2}-u_{1}}{u_{3}-u_{1}}}$. Its general
solution is
\begin{equation}
y = sn\left(x+\delta\right)
\end{equation}
where $\delta$ is an arbitrary constant. Hence for both
connections yielding orbits with perihelia,
\begin{equation}
u-u_{1}=\left(u_{2}-u_{1} \right)sn^{2}
\left(\frac{1}{2}\varphi\sqrt{L_{3}(u_{3}-u_{1})} + \delta
\right) .
\end{equation}
The periodicity of these solutions enables one to calculate a
perihelion shift per revolution. The increase in $\varphi$ between
successive perihelia is  given precisely by
\begin{equation}
\Delta \varphi
=2\int_{u_{1}}^{u_{2}}\frac{du}{\sqrt{L_{3}(u-u_{1})(u-u_{2})(u-u_{3})}}
.
\end{equation}
With the transformation $y=\sqrt{\frac{u-u_{1}}{u_{2}-u_{1}}}$,
this becomes
\begin{equation}
\Delta \varphi =\frac{4K}{\sqrt{(u_{3}-u_{1})L_{3}}}
\end{equation}
where
$$
K = \int_{0}^{1}\frac{dy}{\sqrt{(1-y^{2})(1-p^{2}y^{2})}} .
$$

Depending on circumstances one may be able to approximate this
integral. This is possible if one is interested in
non-relativistic bounded orbits in which the dimensionless
quantity $Mu$ remains small compared with unity and the orbital
speed is small compared with the speed of light. This would, for
example, arise for the motion of the planet Mercury regarded as a
test particle in orbit about the Sun as a source. Even at the
Sun's surface, where $R_{\odot}=7 \times 10^{8} m$,
$M_{\odot}=1.477 \times 10^{3} m$
%(in geometrized units $G=1,c=1$)
, $Mu=2.11 \times 10^{-6}$. At aphelion and perihelion, both
$-\frac{L_{3}}{L_{2}}u_{1}$ and $-\frac{L_{3}}{L_{2}}u_{2}$ are
small quantities \cite{synge}. Thus in the following we
approximate $-\frac{L_{3}}{L_{2}}u_{3}\simeq 1$. This means
$p^{2}$ can be considered  small so:
$$
K \simeq \frac{1}{2} \pi \left(1+\frac{1}{4} p^{2}\right)
$$
and since the ratios $\frac{u_{1}}{u_{3}}$ and
$\frac{u_{2}}{u_{3}}$ are  small, we expand
$$
p^{2} \simeq \frac{(u_{2}-u_{1})}{u_{3}} \simeq -
\frac{(u_{2}-u_{1})L_{3}}{L_{2}}.
$$
We  further approximate the term
$$
\frac{1}{\sqrt{(u_{3}-u_{1})L_{3}}} =
\frac{1}{\sqrt{|L_{2}|\left(1+\frac{L_{3}}{L_{2}}(2u_{1}+u_{2})\right)}}
\simeq
\frac{1}{\sqrt{|L_{2}|}}\left(1-\frac{L_{3}}{2L_{2}}(2u_{1}+u_{2})\right)
$$
After a little
algebra one finds that the increase in $\varphi$ per revolution
becomes
\begin{equation}
\Delta \varphi \simeq
\frac{2\pi}{\sqrt{|L_{2}|}}\left(1-\frac{3L_{3}}{4L_{2}}
 (u_{1}+u_{2})\right).
\label{EB}
\end{equation}
The  advance of perihelion (perihelion shift) per revolution
would be
$$
\varepsilon = \Delta \varphi - 2 \pi.
$$
It is conventional to define   $e$, the eccentricity of an
elliptic orbit and  $r_{0}$ its semi-axis major. Then
$$
r_{1}=(1+e)r_{0}, \qquad r_{2}=(1-e)r_{0}
$$
where $r_{1}=\frac{1}{u_{1}}$ corresponds to aphelion distance and
$r_{2}=\frac{1}{u_{2}}$ corresponds to perihelion distance of an
elliptic orbit. The perihelion shift may be expressed in terms of
these orbit parameters for the limiting Newtonian Kepler ellipse.
Once a set of Kepler orbit parameters have been ascertained then
these formulae permit one to match them to a relativistic orbit
 in terms of  $M$, $l$, $\omega$ and $A$ and the constants
of motion \cite{dereli2}.

Thus in summary, we have examined the bounded orbits of test
particles in the presence of a particular solution to the
Brans-Dicke theory of gravitation. Even with the same constants of
motion and the same limiting Kepler orbits such orbits will in
general differ. Those with  histories determined by the
Levi-Civita connection and satisfying the approximations above,
have a  perihelion shift per revolution
\begin{equation}
\varepsilon_{1} = \frac{2 \pi}{\sqrt{|S_{2}|}}\left(1 +
\frac{3}{2(1-e^2)r_{0}} \frac{S_{3}}{|S_{2}|} \right ) - 2 \pi
.\label{GG}
\end{equation}
By contrast if the history is determined by the connection with
torsion, the perihelion shift per revolution is different, being
given by
\begin{equation}
\varepsilon_2 = \frac{2 \pi}{\sqrt{|C_{2}|}}\left(1 + \frac{3}{2(1-e^{2})r_{0}}
\frac{C_{3}}{|C_{2}|}\right) - 2 \pi .   \label{EG}
\end{equation}

Finally we  note that, when $A=0$ (i.e. the scalar field $\Phi$ is
constant) both orbit equations describe geodesic motion given by
the Levi-Civita connection in a background Kerr geometry. If one
further sets  $l=0$, they both describe geodesic motion in a
background  Schwarzschild geometry. In this  case the constants
reduce to $C_{3}=S_{3}=2M$ and $C_{2}=S_{2}=-1$ and the perihelion
shift reduces to the classical value \cite{synge}

\begin{equation}
\varepsilon =  \frac{6 \pi M}{(1-e^{2})r_{0}} .
\end{equation}

\bigskip
\section{Conclusion}

\bigskip

An analysis of  bounded orbital motion of a massive spinless test
particle in the background of a Kerr Brans-Dicke geometry has been
given. Such orbits  have been discussed in terms of worldlines
that are auto-parallels of different metric compatible  space-time
connections. In one case the connection is torsion-free while in
the other, the connection has a torsion that depends on the
(spacetime) gradient of the scalar field. Such a connection
differs from that of Levi-Civita  by terms that contribute to the
so called ``improved stress-energy tensor" that enters into the
original formulation of the Brans-Dicke theory. The variational
derivation of this theory in terms of independent metric and
connection variations naturally gives rise to a connection with
such torsion.

It is notoriously difficult to discriminate between gravitational
theories by observing planetary orbits in the solar system since
many perturbations need to be taken into account. In the above
calculations one has some information about the value of the
constant $\omega$ but little guidance as to the value of the constants
$\Phi_0$ and  $A$. Furthermore there is no unique Kerr Brans-Dicke
geometry that one may, with confidence, ascribe to any particular
spinning source. Despite these shortcomings, we feel that there
may be astrophysical situations where considerations relevant to
non-Riemannian geometries become relevant. The motion of bodies
around spinning black holes, neutron  or magnetic stars are
examples. In order to obtain evidence for {\it any} gravitational
interaction one needs to analyse observational data in terms of a
particular theoretical framework. We have suggested that the
detection of scalar-tensor gravitational interactions may benefit
from a non-Riemmannian description. The calculations presented
here offer such a framework in the context of the Brans-Dicke
geometry with and without torsion.
The effects of coupling other types of fields such as a Kalb-Ramond 2-form potential
maybe worth investigating \cite{kar}.

\bigskip

\noindent {\large {\bf \centerline{Acknowledgement}}}

\bigskip

\noindent HC thanks the Department of Physics, Lancaster
University, UK  for hospitality. His stay there was made possible
by a Fellowship from  T\"{U}B\.{I}TAK (The Scientific and
Technical Research Council of Turkey) within the BAYG-BDP program.
RWT is grateful to BAE-Systems (Warton) for support for this work.


\newpage
{\small
\begin{thebibliography}{99}

\bibitem{will} C M Will, {\bf Theory and Experiment in Gravitational Physics}
(Cambridge U P , 1993).

\bibitem{will2}  C M Will, {\it The Confrontation between General
Relativity and Experiment}, gr-qc/0103036 .

\bibitem{pulsar} A R Hulse, J H Taylor, Ap. J. Lett. {\bf 195} L51 (1975).

\bibitem{einstein} A Einstein, L Infeld, B Hoffmann, Ann. Math. {\bf
39} 65  (1938).

\bibitem{papapetrou} A Papapetrou, Proc. Roy. Soc. {\bf A209} 248 (1951).

\bibitem{dixon} W G Dixon, Proc. Roy. Soc. {\bf A314} 499 (1970).

\bibitem{weinberg} S Weinberg, {\bf Gravitation and Cosmology} ( Wiley, 1972).

\bibitem{brans} C. H. Brans, R. Dicke, Phys. Rev. {\bf 124}(1961)925.

\bibitem{brans97} C. H. Brans, {\sl Gravity and ther tenacious scalar fields} (gr-qc/9705069)
Contribution to Engelbert Sch\"{u}cking Festschrift.

\bibitem{Dirac} P A M Dirac, Proc. Roy. Soc. {\bf A333} 403 (1973).

\bibitem{DT} T Dereli, R W Tucker, Phys. Letts. {\bf B110} 206 (1982).

\bibitem{dereli1} T. Dereli, R. W. Tucker, {\it On the motion of matter in
spacetime}, gr-qc/0107017.

\bibitem{dereli2} T. Dereli, R. W. Tucker, Mod. Phys. Lett. {\bf A17}
(2002)421  (gr-qc/0104050).

\bibitem{intosh} C. B. G. McIntosh, Comm. Math. Physics {\bf 37}(1974) 335.

\bibitem{kim} H. Kim, Phys. Rev. {\bf D60} (1999) 024001.

\bibitem{schmidt} W. Schmidt, Class. Q. Grav. {\bf 19} (2002) 2743.

\bibitem{wilkins} D. C. Wilkins, Phys. Rev. {\bf D5} (1972) 814.

\bibitem{synge} J. L. Synge, {\bf Relativity: The General Theory}
(North-Holland Publishing, 1960).

\bibitem{kar} S. Kar, S. SenGupta, S. Sur, Phys. Rev. {\bf D67} (2003) 044005.

\end{thebibliography}}
\end{document}